\documentclass[journal,twoside,web]{ieeecolor}
\usepackage{jsen}
\usepackage{cite}
\usepackage{amsmath,amssymb,amsfonts}
\usepackage{algorithmic}
\usepackage{graphicx}
\usepackage{color}
\usepackage{textcomp}
\usepackage{wrapfig}
\usepackage{units}
\usepackage{subfig}
\usepackage[utf8]{inputenc}
\usepackage{epstopdf}
\usepackage{caption}
\usepackage{booktabs}
\usepackage{multicol}
\usepackage{acro}
\usepackage{url}
\graphicspath{{figures/}}
\DeclareAcronym{lri}{
  short = LRI ,
  long  = Laser Ranging Interferometer ,
  sort  = A ,
}

\DeclareAcronym{kbr}{
  short = KBR ,
  long  = K-Band Ranging ,
  sort  = A ,
}

\DeclareAcronym{fsm}{
  short = FSM,
  long  = Fast Steering Mirror ,
  sort  = A ,
}

\DeclareAcronym{lsm}{
  short = LSM ,
  long  = Laser Steering Mirror,
  sort  = A ,
}

\DeclareAcronym{pss}{
  short = PSS ,
  long  = Position Sensing System ,
  sort  = A ,
}

\DeclareAcronym{imu}{
  short = IMU ,
  long  = Inertial Measurement Unit ,
  sort  = A ,
}

\DeclareAcronym{los}{
  short = LOS ,
  long  = Line-Of-Sight ,
  sort  = A ,
}

\DeclareAcronym{lriof}{
  short = LRIoF ,
  long  = LRI optical frame ,
  sort  = A ,
}

\DeclareAcronym{srf}{
  short = SRF ,
  long  = Science Reference Frame ,
  sort  = A ,
}

\DeclareAcronym{dws}{
  short = DWS ,
  long  = Differential Wavefront Sensing ,
  sort  = A ,
}

\DeclareAcronym{aocs}{
  short = AOCS ,
  long  = Attitude and Orbit Control System ,
  sort  = A ,
}

\DeclareAcronym{gps}{
  short = GPS ,
  long  = Global Positioning System ,
  sort  = A ,
}
 
 \DeclareAcronym{chu}{
  short = CHU ,
  long  = Camera Head Unit ,
  sort  = A ,
}

\DeclareAcronym{lrp}{
  short = LRP ,
  long  = Laser Ranging Processor ,
  sort  = A ,
}
\DeclareAcronym{oba}{
  short = OBA ,
  long  = Optical Bench Assembly ,
  sort  = A ,
}
\def\BibTeX{{\rm B\kern-.05em{\sc i\kern-.025em b}\kern-.0Fem
   T\kern-.1667em\lower.7ex\hbox{E}\kern-.125emX}}
\definecolor{abstractbg}{rgb}{0.89804,0.94510,0.83137}
\begin{document}
\title{Analysis of GRACE Follow-On Laser Ranging Interferometer derived inter-satellite pointing angles}
\author{Sujata Goswami, Samuel P. Francis, Tamara Bandikova and Robert E. Spero 
\thanks{\copyright 2021 California Institute of Technology. Government sponsorship acknowledged. Accepted for publication. Citation information: DOI 10.1109/JSEN.2021.3090790}
\thanks{Sujata Goswami, Samuel P. Francis and Robert E. Spero are with the NASA Jet Propulsion Laboratory (JPL), California Institute of Technology, 4800 Oak Grove Dr, Pasadena, 91109, California  (e-mail: sujata.goswami@jpl.nasa.gov; samuel.p.francis@jpl.nasa.gov; respero@jpl.nasa.gov). }
\thanks{Tamara Bandikova (e-mail: bandikova.tamara@gmail.com).}}
\IEEEtitleabstractindextext{%
\fcolorbox{abstractbg}{abstractbg}{%

\begin{minipage}{\textwidth}%
\begin{abstract}
	Gravity Recovery and Climate Experiment Follow-On (GRACE-FO) was launched on May 22, 2018. It carries the Laser Ranging Interferometer  (LRI) as a technology demonstrator that measures the inter-satellite range with nanometer precision using a laser-link between satellites. To maintain the laser-link between satellites, the LRI uses the beam steering method: a Fast Steering Mirror  (FSM) is actuated to correct for misalignment between the incoming and outgoing laser beams.
	From the FSM commands, we can compute the inter-satellite pitch and yaw angles. These angles provide information about the spacecraft's relative orientation with respect to line-of-sight (LOS).
	We analyze LRI derived inter-satellite pointing angles for 2019 and 2020. Further, we present its comparison with the pointing angles derived from GRACE-FO SCA1B data, which represents the spacecraft attitude computed from star cameras  and Inertial Measurement Unit (IMU) data using a Kalman filter. We discuss the correlations seen between the laser based attitude data and the spacecraft temperature variations. 
	This analysis serves as the basis to explore the potential of this new attitude product obtained from the Differential Wavefront Sensing  (DWS) control of a FSM. 
\end{abstract}

\end{minipage}}}
% ---
\maketitle
\printacronyms  

\section{Introduction}
	Gravity Recovery and Climate Experiment Follow-On (GRACE-FO), \cite{gfo2019,felix2020} a successor of the GRACE mission \cite{tapley2004,tapley2019}, was launched on May 22, 2018. The satellites measure the inter-satellite biased range   using the \ac{kbr} microwave instrument. These inter-satellite range observations are used to compute the gravity field variations in the Earth system. The GRACE-FO mission carries the \ac{lri} as a technology demonstrator \cite{sheard2012}. \Ac{lri} provides the inter-satellite ranging measurements similar to microwave instrument, but with the precision up to $\unit[1]{nm \; {Hz}^{-1/2}}$ in frequencies above $\unit[100]{mHz}$ \cite{abich2019,gfo2019}. 
	The GRACE-FO spacecraft attitude is measured by the star cameras and \ac{imu}, and is controlled by magnetic torquers and attitude control thrusters.  The precise attitude determination and control is necessary to maintain the microwave instrument inter-satellite pointing within $\unit[2.5]{mrad}$ in roll and $\unit[250]{\mu rad}$ for pitch and yaw.

The \Ac{lri} ranging measurement is performed by interfering the laser on the local satellite with the received light from the laser transmitted by the far satellite, and tracking the phase of the resulting beatnote. To maximize the carrier-to-noise density of the beatnote, the misalignment between local and received beams needs to be minimized.
Precise laser beam pointing is achieved by using a two-axis \ac{fsm}. The actuation angles are derived from \ac{dws} \cite{daniel2014,abich2019}. 
   An eddy-current based \ac{pss} encodes the mirror actuation angles relative to the spacecraft frame.  We analyze the \Ac{pss} signals as they are the \Ac{lri}-based readout of spacecraft orientation. The \Ac{pss} determines spacecraft pitch and yaw, no roll information is provided by \Ac{dws}.

The inter-satellite pitch and yaw angles provided by \Ac{dws} are independent of the \ac{aocs} and represent the spacecraft's relative orientation. They are  published as the GRACE-FO LSM1B data product.
The primary information about absolute and relative orientation is provided by the three star cameras and \Ac{imu} (consisting of four fiber optic gyroscopes). GRACE-FO SCA1B data represents the attitude solution, obtained by the combination of  star camera and \Ac{imu} data using a Kalman filter. 
The comparison of LSM1B and SCA1B data will provide information about the mutual consistency  of  these two products. It will also help us to characterize any potential systematics or errors in the individual sensor measurements. Further, having understood the signal and noise characteristics of each source, the LSM1B data can be used to improve the attitude solution.  For gravity field determination, precise information about spacecraft's attitude is needed to \cite{bandi2014} 
	\begin{enumerate}
		\item Compute the antenna phase center offset correction for K-band ranging observations
		 \item Rotate the accelerometer data from science reference frame to Earth-centered inertial reference frame 
		 \item Rotate the \ac{gps} phase center from science reference frame to Earth-centered inertial reference frame
	\end{enumerate}
Any error in the SCA1B data will  propagate into the gravity field models. Thus, the reduction of attitude noise using LSM1B promises to improve gravity field models. 

The attitude measurements can also potentially be used to reduce noise in the \Ac{lri} range measurements. Micrometer-level offsets between the triple mirror assembly vertex and the satellite center of mass results in spacecraft attitude variations coupling into the LRI range measurement. This is an example of tilt-to-length coupling error. 
To reduce the error from tilt-to-length, the attitude measurements can be combined with coupling coefficients, which have been estimated from the center-of-mass maneuvers performed by the GRACE-FO satellites \cite{henry2020}.
	
Here we present the first analysis of the laser beam pointing data and provide details about the processing itself. Understanding of this new dataset is a key factor for further improvement of the spacecraft's attitude solution and consequently for improvement of the gravity field models derived from K-band ranging and laser ranging observations.
 
 In Sec. \ref{sec_beamsteering} we explain how the GRACE-FO \ac{lri} is able to provide measurements of spacecraft attitude.  In Sec. \ref{sec:compare_dws_sca} we compare these laser attitude measurements with the primary GRACE-FO attitude data product, SCA1B. Finally, in Sec. \ref{sec_outlook} we discuss the findings of this comparison and describe how the laser attitude measurements could be incorporated into the primary GRACE-FO attitude data product.
%%%	
\section{LRI-based readout of inter-satellite pointing} \label{sec_beamsteering}
\begin{figure}[htbp!]
	    \centering
	    \includegraphics[width=0.70\textwidth]{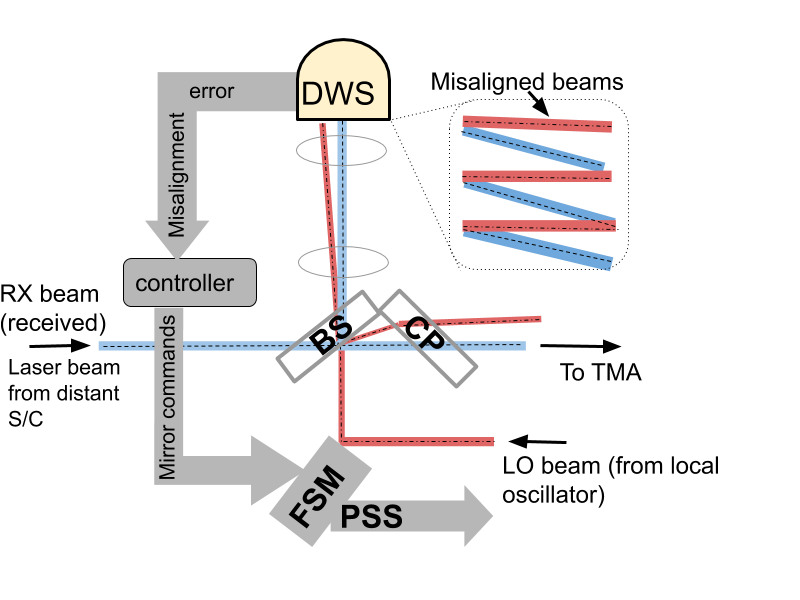} \\
	    \caption{Steering mirror control loop zeroing the \Ac{dws} signal by rotating the steering mirror such that Local       Oscillator  (LO) and incoming or received beam (RX) are co-aligned. FSM: fast steering mirror, DWS: Differential wavefront sensing, BS: beam splitter, CP: compensation plate, TMA: Triple mirror assembly, PSS: Position sensing system, S/C: Spacecraft. }
	    \label{fig:FSM_function}
	\end{figure}

	As described in \cite{sheard2012,daniel2014}, the LRI senses the
	pointing of the outgoing  or Local Oscillator (LO) beam relative to the incoming beam or Receiver (RX) beam from the distant spacecraft and controls the outgoing beam direction so the two
	beams overlap (see Fig. \ref{fig:FSM_function}).  The angle sensing, or \Ac{dws}, is done by subtracting the
	phases from opposite halves, vertically and horizontally, across
	quadrant photodiodes  (The sum phase measures the along-track
	spacecraft separation).  A control system actuates the  \Ac{fsm} with high
	gain to keep the  \Ac{dws} error signal nulled.  \Ac{pss} encoders measure  \Ac{fsm}
	actuation angles relative to the spacecraft frame.  Together,  \Ac{dws},
	control system, and  \Ac{fsm} act as an active retroreflector, aligning the
	outgoing beam direction to overlap the incoming beam.  

	The PSS signals provide a readout of the \ac{los} angles to the
	distant spacecraft.  In the absence of relative transverse motion
	between the spacecraft, the \Ac{pss} signals measure changes in the
	pitch and yaw of the spacecraft.  \Ac{dws} does not provide roll information.
	Inter-satellite pointing angles can also be derived from \Ac{aocs} information provided by the star camera and \Ac{imu} sensors. This allows us to compare the spacecraft pointing by \Ac{dws} vs SCA1B data. 

%% ~~~~~~~~~~~~~~~~~~~~~~~~~~~~~~~~
\subsection{Computation of steering mirror pointing angles from the steering mirror commands}\label{sec_1a21b}
    There are two ways to calculate \Ac{los} angles from \Ac{lri} telemetry: using either mirror commands ($\text{C}_x$ and $\text{C}_y$), or using the sensed orientation of the mirror ($\text{S}_x$ and $\text{S}_y$). As Fig. \ref{fig:FSM_function} shows, mirror commands from the \Ac{dws} loop controller are sent to \Ac{fsm} to change the mirror tip/tilt in response to a misalignment between local and received beams. The geometry of the \Ac{lri} optical bench determines how these mirror commands map to \Ac{los} pitch and yaw angles in the \ac{lriof}. This mapping is described in \eqref{eq:1a_1b}.  The   tip/tilt of the  \Ac{fsm} in response to these mirror commands is sensed internally by the \Ac{pss}. Equation \eqref{eq:s_xy} shows the mapping from the sensed measurements ($\text{S}_x$ and $\text{S}_y$) into  \Ac{los} pitch and yaw angles in the \Ac{lriof}.
    
Both mirror commands and the sensed mirror orientation are included in the LSM1A data product. \Ac{pss}  data are used to derive the \Ac{lri} \Ac{los} pitch and yaw angles provided in the LSM1B data.

 From the mirror commands ($\text{C}_x$ and $\text{C}_y$) in LSM1A dataset, the inter-satellite pointing pitch  $(\phi)$ and yaw  $(\theta)$ angles are computed as  
    %           LSM1A to LSM1B processing
	   \begin{equation} \label{eq:1a_1b}
			\begin{bmatrix}
				\theta \\ \phi
				\end{bmatrix} = 
				\text{R}
				\begin{bmatrix}
				\text{C}_x \\ \text{C}_y
				\end{bmatrix} \kappa 
		\end{equation} 
		in micro radians ($\unit[]{\mu rad}$). 
	
	From the \Ac{pss} \textit{counts} ($\text{S}_x$ and $\text{S}_y$), the inter-satellite pointing pitch  $(\phi)$ and yaw  $(\theta)$ angles are computed as  
			  \begin{equation} \label{eq:s_xy}
				\begin{bmatrix}
				\theta \\ \phi
				\end{bmatrix}  = 
				\text{R}
				\begin{bmatrix}
					  a_x\text{S}_x + b_x \\    a_y\text{S}_y + b_y
				\end{bmatrix}  \kappa.	
	   	\end{equation}
	In Equation \eqref{eq:1a_1b} \& \eqref{eq:s_xy}, matrix $\text{R}$, rotating the observations from instrument frame to \Ac{lriof}, is given as
        	         \begin{equation}\label{eq:frame_rot}
                   	\text{R} =
                            			\begin{bmatrix}
          							 -1 & 1  \\
           							 1/\sqrt{2} & 1/\sqrt{2}
          					\end{bmatrix}.
                \end{equation}
 		$\kappa = \unit[4.044]{\mu rad \;count^{-1}}$  in  \eqref{eq:1a_1b} \& \eqref{eq:s_xy} defines the measurement precision of the  \textit{counts} which are recorded from the FSM. 
		The coefficients $a_x, a_y, b_x$  and  $b_y$  in  \eqref{eq:s_xy} are given in the Table \ref{table1}. The values of the coefficients were measured on ground. During on-ground tests before mission launch, the steering mirrors were commanded using different patterns and these values were determined by fitting the commands to the sensed \Ac{pss} values. So ($a, b$) values map between mirror command and sensed \Ac{pss} angles. In Table \ref{table1}, GRACE-FO 1 and 2 are represented as GF1 and GF2 respectively. This notation is used throughout this paper to refer the two spacecrafts.
		   			\begin{table}[htbp!]  
							\caption{Coefficients used in \eqref{eq:s_xy} to compute the PSS  \textit{counts} for the two spacecraft. }
								\begin{center}
									\begin{tabular}{|c|c|c|c|c|}
										SC 	& 	$a_x$ 	& 	$a_y$ 	& 	$b_x$ 	& 	$b_y$ \\
										\hline \\
										GF1 & 	1.130289	&	1.131806	&	-2064.4	&	-2068.1 \\
										GF2 &	1.131136	&	1.131948	&	-2064.8	&	-2068.2 \\
									\end{tabular}
								\end{center}
							\label{table1}
					\end{table}%

		\begin{figure}[htbp!]
		\centering
			\includegraphics[width=0.53\textwidth]{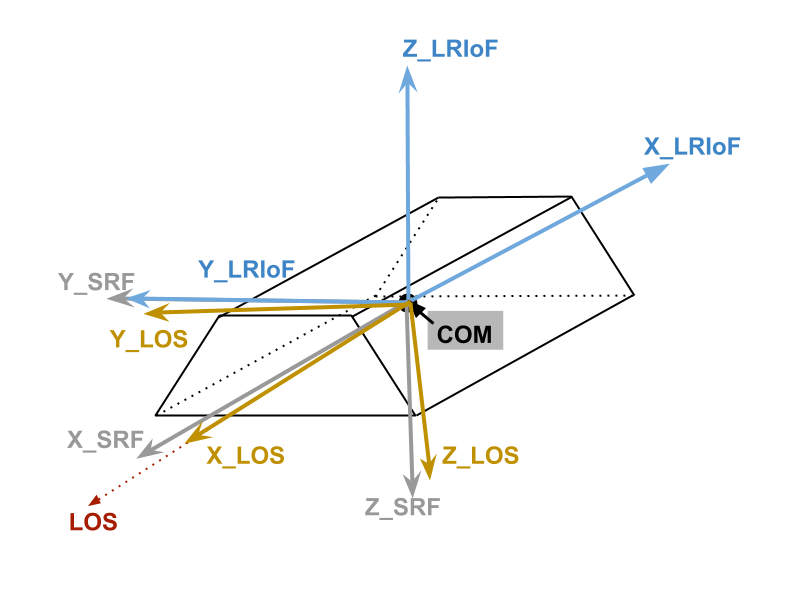}
			\caption{Accomodation of the \ac{lriof},  the \ac{srf} and the \ac{los} frame in the GRACE-FO spacecraft. COM is the center-of-mass of the spacecraft.}
			\label{fig:frame}
		\end{figure}		
		
		Afterwards, the computed angles are time-tag corrected to get the angles corresponding to GRACE-FO \Ac{gps} time. Then, they are rotated from the \Ac{lriof} to the \Ac{srf}.  The relation between \Ac{lriof} and \Ac{srf} is given as (cf. Fig. \ref{fig:frame})
		 \begin{equation}
			\begin{bmatrix}
			    X_\textsc{srf} \\ Y_\textsc{srf} \\ Z_\textsc{srf}
			\end{bmatrix} = \begin{bmatrix}
				                   -X_\text{LRIoF} \\ Y_\text{LRIoF} \\ -Z_\text{LRIoF}
				                \end{bmatrix} 
			\label{eq:srf_lri}
		\end{equation}
	These inter-satellite pointing angles are provided in the LSM1B product by JPL and represent variations in \Ac{srf} with respect to \Ac{los} (see Fig. \ref{fig:frame}).
 % ~~~~~~~~~~~~~~~~~~~~~~~
\section{A comparison between the LSM1B and SCA1B GRACE-FO attitude products}\label{sec:compare_dws_sca}
  In this section we compare the laser attitude measurements, LSM1B, with the estimated attitude SCA1B which is a  combination of star camera head units and \Ac{imu} measurements. This comparison serves two purposes: (1) it allows for an independent verification of the SCA1B data product; and (2) since GRACE-FO is the first inter-spacecraft laser interferometer, it provides an opportunity to evaluate the performance of the \ac{dws} technique.
   
   SCA1B data is computed by on-ground fusion of star cameras  and \Ac{imu} data, in an attitude Kalman filter \cite{harvey2019,bandi2018}. The Kalman filter used for GRACE is adapted to GRACE-FO. The GRACE SCA1B attitude estimation used  two star cameras and angular accelerations measurements from the accelerometer as input to the Kalman filter \cite{harvey2019}, and the SCA1B attitude estimation for GRACE-FO uses three star cameras and an \Ac{imu}  data as input to the Kalman filter. It is expected that with three star cameras, periods of complete star camera blinding will be less frequent. The location of the three star cameras on-board, one on the top and two are on each of the lateral sides of the spacecraft, and \Ac{imu} (which consists of four fibre optic gyroscopes) is shown in Fig.\ref{fig:spacecraft_design}.  All the available attitude observations at an epoch are combined to compute the SCA1B product.
        
  Section \ref{sec_1a21b} explained the LSM1B angles represent variations in between the \Ac{los}  and \Ac{srf},  to allow a comparison between LSM1B and SCA1B we derive  similar pointing angles representation from the SCA1B product as well. 
    SCA1B data contain quaternions representing the spacecraft attitude in the Earth-centered inertial frame of reference in terms of quaternions ($q ={q_0, q_1, q_2, q_3}$). From these quaternions, the roll, pitch and yaw angles are computed in the \Ac{srf} with respect to \ac{los}  using the spacecraft orbit in Earth-centered inertial  frame, as explained by \cite{bandi2012}. In the following, we use ``SCA1B$_\text{\ac{los}}$" designation for the pointing angles derived from SCA1B data. 

 The LSM1B angle measurements are independent of the star camera and \ac{imu} measurements used to generate the SCA1B$_\text{\ac{los}}$ angles. Comparing LSM1B and SCA1B$_\text{\ac{los}}$ angles in the SRF  provides  a way to validate the Kalman filter based SCA1B dataset. Since the SCA1B product is used in the gravity field determination from K-band ranging observations, it is one of the many factors contributing to the overall precision of the gravity field solutions. Hence an improved SCA1B product will enhance the precision of gravity field model. 
  \subsection{Differences  between SCA1B$_\text{\ac{los}}$ and LSM1B angles in time domain} \label{sec:temporal}
   In the year 2018, right after the GRACE-FO was launched, in July, one of the instrument processing units  \cite{abich2019} onboard failed, making \ac{lri} data unavailable from mid-July to mid-December, 2018 \cite{gfoupdate}.   After 2018,  \ac{lri} data was available for most of the time. Hence, we consider the analysis of LSM1B data for the year 2019 and 2020.  
   
   In this section, we present the comparison between SCA1B$_\text{\ac{los}}$ and LSM1B angles, at first over a day (in Sec. \ref{sec:day}).  Further, a comparison between the long-term time-series is presented in Sec. \ref{sec:long-term} and \ref{sec:longterm}.
   	\subsubsection{Difference between SCA1B$_\text{\ac{los}}$ and LSM1B over a day}\label{sec:day}
\begin{figure*}[htbp!] % python ... .testscripts/lsmsca.py
		 	\centering
		 		\includegraphics[width=0.95\textwidth, trim={27 0  50 30}, clip]{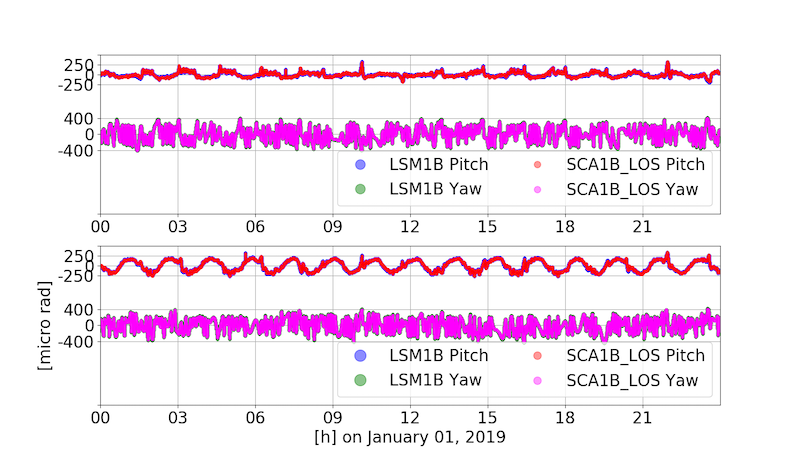} 	 	
			\caption{Pitch and yaw angles of GF1 (top panel) and GF2 (bottom panel) as observed by the LSM1B and SCA1B$_\text{\ac{los}}$. LSM1B data is resampled from 10s to 1s. The two set of angles are completely overlapped in this plot.}
			\label{py_cd}
	\end{figure*}

		In Fig. \ref{py_cd}, we have plotted the  LSM1B and SCA1B$_\text{\ac{los}}$ angles for January 01, 2019, for GF1 and GF2, both.  The angles in LSM1B data have a bias, a result of arbitrary manufacturing offsets in the \ac{fsm}. The LSM1B data shown in this paper have this bias removed. The continuous time-series of LSM1B pitch and yaw angles shows the changes to the \ac{fsm} orientation needed to keep the LRI laser beams aligned.	
		 \begin{figure*}[htbp!]
		 	\centering	
				\includegraphics[width=0.95\textwidth, trim={24 0  40 27}, clip]{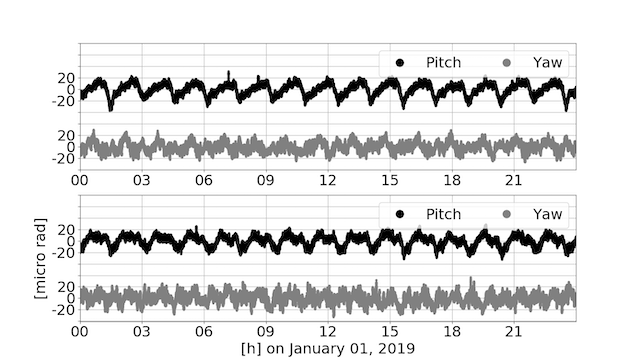}						
				\caption{Differences in the pointing angles between LSM1B and SCA1B$_\text{\ac{los}}$. The differences are shown for GF1 (top panel) and GF2 (bottom panel), both.}
			\label{py_cd_diff}
		\end{figure*}
  The differences in two sets of the inter-satellite pointing angles vary  up to $\unit[40]{\mu \text{rad}}$ peak-to-peak (as shown in Fig. \ref{py_cd_diff}) which are at the required precision level of the spacecraft attitude.  
       
   The pitch differences shown in Fig. \ref{py_cd_diff} are dominated by signals with a frequency of once-per-rev. The yaw differences are dominated by both once-per-rev and twice per-rev signals. The signals at the orbital period ($\approx$93 minutes) suggests this difference is driven by temperature changes \cite{save2019}. Although we have been unable to conclude how this temperature dependence enters into the LSM1B and SCA1B measurements,   possible explanations are discussed in Section \ref{sec:long-term} and \ref{sec:longterm}.
%~~~~~~~~~~~~~~~~~~~~~~~~~~~~~~`
\subsubsection{Analysis of long-term characteristics of the two sets of inter-satellite pointing angles } \label{sec:long-term}
In order to understand the differences between the pointing angles derived from SCA1B and LSM1B data, we analyze their long-term characteristics. In Figs.\ref{fig:py_lrp_gf1} \& \ref{fig:py_lrp_gf2}, the daily mean values of inter-satellite pointing angles are shown for the year 2019 and 2020. GRACE-FO \ac{aocs}, which comprises star cameras, \ac{imu}, thrusters and magnetic torquers, keeps pitch and yaw variations of both spacecraft within $\pm \unit[100]{\mu rad}$ over time scales slower than the orbital period as seen in Figs.\ref{fig:py_lrp_gf1} \& \ref{fig:py_lrp_gf2}. 
During February 06 to March 17, 2019 there were reset problems on the instrument processing unit of GF2 and during April 24 to April 29, 2019, the \ac{lri} was in diagnostic mode. Hence, no  \ac{lri} data was available during these periods.   From January 18 to 23, 2020  \ac{kbr} and  \ac{lri}  link was lost after GF2 switched to  \ac{aocs} safe mode. During Feb 7 to 13,  2020 the  \ac{kbr} and  \ac{lri} measurements were interrupted. Hence, no  \ac{lri} data was available during these periods  (see Figs.\ref{fig:py_lrp_gf1} \& \ref{fig:py_lrp_gf2}).

During center-of-mass calibration maneuvers   (marked as (ii) in Figs.\ref{fig:py_lrp_gf1} \& \ref{fig:py_lrp_gf2}), which in the first mission phase occurred monthly,  the LSM1B pitch angles on both spacecraft reached up to $\unit[1.2]{} - \unit[2] {mrad}$, and, LSM1B yaw angles reached up to $\unit[4] {mrad}$. 
The center-of-mass calibration maneuvers show that the  \ac{dws}/ \ac{fsm} control system is able to  continuously track attitude changes up to a few mrads while maintaining overlap between the local and received laser beams. The threshold of this tracking was tested on July 19, 2018 when an error caused the GF2  \ac{aocs} to transition into attitude safe mode. When this happened, the  \ac{lri} was able to maintain a science link, tracking the change in spacecraft attitude, until the attitude exceeded $\unit[6.5] {mrad}$. The limit was the light traveling between the spacecraft hitting the light baffle and no light reaching the detector.

At nearly $\beta^{'}=0$ crossings (for example in April and September 2019), \ac{chu} number 1 or CHU1  was blinded which led to the increased errors in SCA1B$_\text{ \ac{los}}$ yaw angles. That’s why we see SCA1B$_\text{LOS}$ yaw daily mean variations are more scattered than the LSM1B yaw angles (features marked as (i) in Fig.\ref{fig:py_lrp_gf1}). 
 
In June, 2020 the satellite  \ac{aocs} was updated to allow smoother transitions between star cameras for the periods where blindings limit the availability of all cameras. As Figs. \ref{fig:py_lrp_gf1} \& \ref{fig:py_lrp_gf2} show, these updates which were performed on the GF1  \ac{aocs} on June 2 and on GF2   \ac{aocs}  on June 17, introduced a bias change in the satellite pointing.

The  daily mean values of the SCA1B$_\text{\ac{los}}$ and LSM1B pointing angles are seen to be correlated with the $\beta^{'}$ angle (cf. Figs.\ref{fig:py_lrp_gf1} \& \ref{fig:temps_gf1} and   \ref{fig:py_lrp_gf2} \& \ref{fig:temps_gf2}).  The $\beta^{'}$  is the angle between the orbital plane and the Earth-Sun vector. It varies between $\pm90^{\circ}$ \cite{montenbruck2002}.  
\begin{figure}[htbp!]
	\includegraphics[width= 0.5\textwidth]{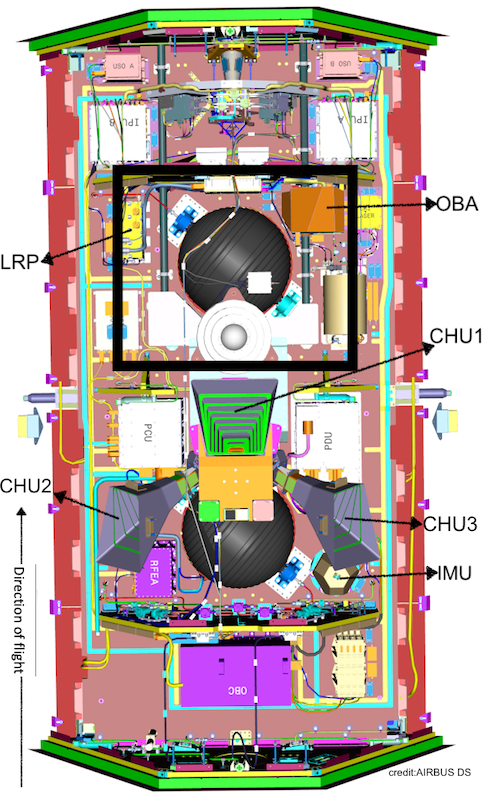}%spacecraft_designdescriptionDoc.png}}
	\caption{Accomodation of sensors on GRACE-FO spacecraft. Attitude sensors: Star camera head unit 1, 2 and 3,   \ac{imu}. The  \ac{lri} components used to perform active beam steering are contained within the black bounded box. The temperature of highlighted sensors: Laser Ranging Processor (LRP), Optical Bench Assembly (OBA), Camera Head Unit1 (CHU1) are shown in Figs. \ref{fig:temps_gf1} \& \ref{fig:temps_gf2}.}
	\label{fig:spacecraft_design}
\end{figure}
 %
%%
% python ... .testscripts/plot_offred_pitchLSM.py
 
 %% ~~~~~~~~~~~~~~~
 \subsubsection{Investigation of SCA1B$_\text{\ac{los}}$ and LSM1B correlations with $\beta^{'}$} \label{sec:longterm}
\begin{figure*}[htbp!]
\centering
\subfloat[Daily mean variations  of pitch and yaw angles  of SCA1B$_\text{\ac{los}}$ and LSM1B  of GF1 during 2019 and 2020. The features marked in box (i) show the scattering in SCA1B$_\text{\ac{los}}$ yaw angles due to increased errors in SCA1B solution during this period; (ii) shows the large angle values due to center-of-mass calibration maneuvers;(iii) shows the change in attitude bias after updating the quaternions on \ac{chu}2 and \ac{chu}3; (iv) shows the attitude affected during thruster calibration tests. The difference of LSM1B and SCA1B$_\text{\ac{los}}$ pitch angles are plotted in Fig. \ref{fig:diff_scaimu}.]{
	\label{fig:py_lrp_gf1}
 	\includegraphics[width= 0.94\textwidth, trim={8 4  24 10}, clip]{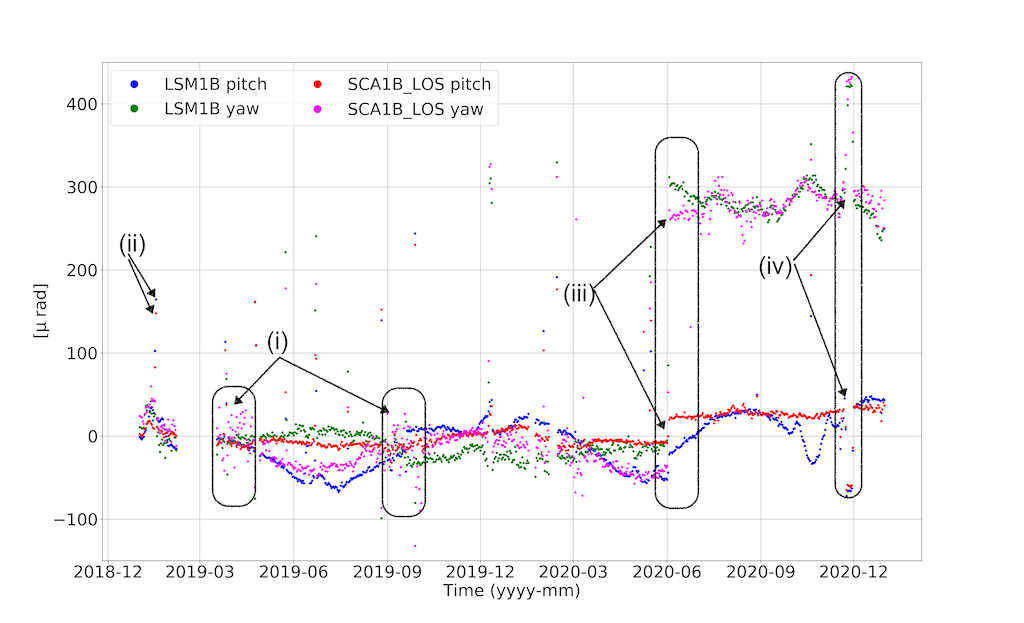}} \\
\subfloat[Relevant  spacecraft temperatures of GF1 vs $\beta^{'}$  angle during 2019 and 2020. LRI LRP: LRI Laser Ranging Processor, LRI OBA: LRI Optical Bench Assembly,  \ac{chu}1: Star Camera Head Unit 1.]{
	\label{fig:temps_gf1}
	\includegraphics[width=0.94\textwidth,trim={12 4  24 10}, clip]{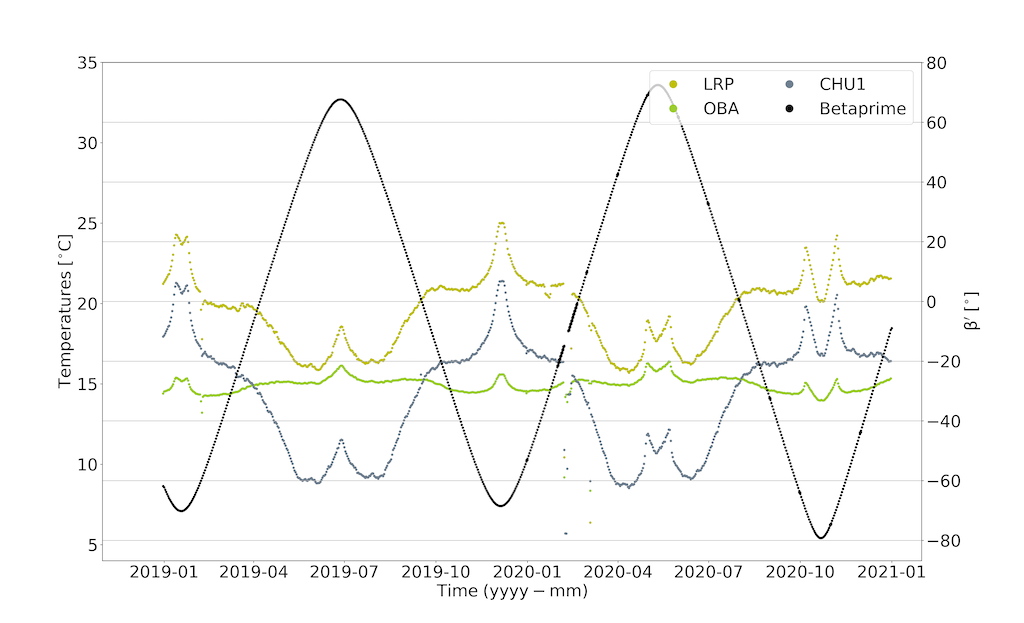} }
\caption{Comparison of Fig. \ref{fig:py_lrp_gf1} \& \ref{fig:temps_gf1}  show the correlations between GF1 angles, $\beta^{'}$ and spacecraft temperatures during 2019 and 2020.  Correlations between the LSM1B pitch angles and the spacecraft temperatures (\ac{lrp}, \ac{chu}1) are also seen in the two figures.}
\label{fig:lrptemps_1}
\end{figure*}

\begin{figure*}[htbp!]
\centering
\subfloat[Daily mean variations  of pitch and yaw angles  of SCA1B$_\text{\ac{los}}$ and LSM1B of GF2 during 2019 and 2020. The features marked in (ii) shows the large angle values due to center-of-mass calibration maneuvers;(iii) shows the change in attitude bias after updating the quaternions on \ac{chu}2 and \ac{chu}3.]{
	\label{fig:py_lrp_gf2}
 	  \includegraphics[width= 0.94\textwidth, trim={8 4  24 10}, clip]{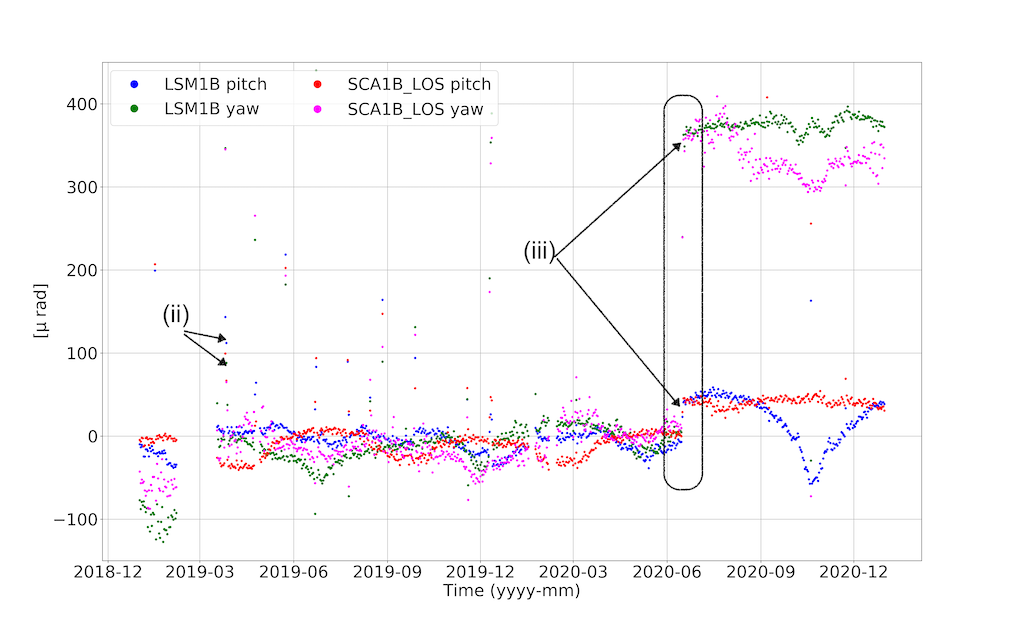}} \\
\subfloat[Relevant  spacecraft temperatures of GF2 vs $\beta^{'}$  angle during 2019 and 2020. LRI LRP: LRI Laser Ranging Processor, LRI OBA: LRI Optical Bench Assembly,  \ac{chu}1: Star Camera Head Unit 1.]{
	\label{fig:temps_gf2}
	\includegraphics[width=0.94\textwidth,trim={8 4  15 6}, clip]{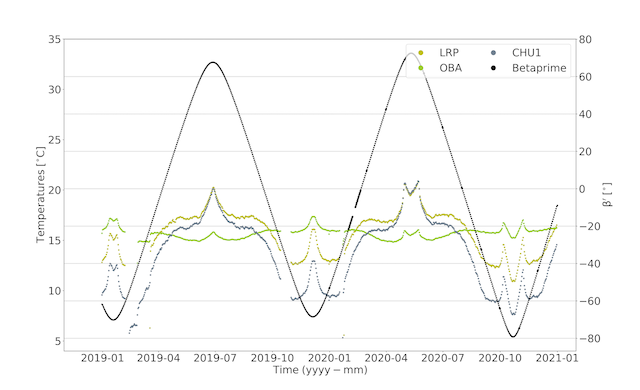} }
\caption{Comparison of Figs.\ref{fig:py_lrp_gf2} \& \ref{fig:temps_gf2} show the correlations between GF2 angles,  $\beta^{'}$ during 2019 and 2020 along with the relevant spacecraft temperatures.}
\label{fig:lrptemps_2}
\end{figure*}

The  correlations with the $\beta^{'}$  indicate a possible thermal dependence since the correlations between the spacecraft temperatures and $\beta^{'}$ variations has been seen in other GRACE-FO data \cite{save2019}. Therefore, we investigated whether the LSM1B and/or SCA1B$_\text{\ac{los}}$ angles correlated with temperature measurements across the spacecraft. We looked for correlations between LSM1B/SCA1B$_\text{LOS}$ angles and temperatures from the \ac{chu}1 and \ac{lri}. Since the \ac{lri} components are spread across the spacecraft, we investigated the temperatures of the individual components: \ac{lrp}, \ac{oba}. The temperatures during 2019 and 2020 are shown for GF1 and GF2  in Figs.\ref{fig:temps_gf1}  \& \ref{fig:temps_gf2}  respectively.  The instruments along with their location are highlighted in the spacecraft in Fig.\ref{fig:spacecraft_design}.

Among the temperatures shown in Figs. \ref{fig:py_lrp_gf1} \& \ref{fig:temps_gf1}, the LSM1B pitch angles (on both GF1 and GF2) show moderate correlations with temperatures across the spacecraft (\ac{lrp}, \ac{chu}1). The LSM1B yaw angles show weak correlation, as did the SCA1B$_\text{\ac{los}}$ pitch and yaw angles. 

We emphasize that the LSM1B and SCA1B measurements are within specification. Since this is the first time a laser interferometer has been used between spacecraft, we are interested in understanding the limits of the LSM1B measurements. We therefore attempted to explain the long-term temperature correlated drift, most prominent in the GF1 LSM1B pitch measurements. We have so far not been successful in explaining this drift but do have several hypotheses that require further investigation.

Although the temperature sensitivity appears strongest in the LSM1B angles, the smaller correlation between SCA1B$_\text{LOS}$ angles and temperature could be because the star camera and \ac{imu} measurements are used to actively correct the satellite attitude. Temperature sensitivities in the star cameras or \ac{imu} would be applied as corrections to the spacecraft pointing via the \ac{aocs} that maintains the satellite attitude. These would then appear in the LSM1B measurements as real satellite attitude fluctuations. We therefore started by investigating whether the temperature sensitivity of any of the attitude sensors could explain the temperature correlated drift in the daily mean LSM1B pitch measurements.

\begin{figure*}[htbp!]
        \centering
        \includegraphics[width=0.96\textwidth, trim={5 0 11 7}, clip]{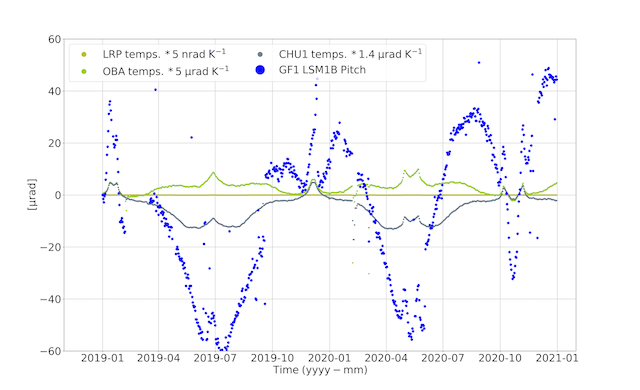}  
        \caption{LSM1B pitch measurements for GF1 are plotted against predictions of the long-term drift due to temperature sensitivities in the \ac{chu}1,  optical bench assembly (OBA) and laser ranging processor (LRP). The temperature coefficient for the star camera head are provided by John Leif J\o rgensen, DTU, Denmark and the temperature coefficients for the  \ac{oba}  and  \ac{lrp},  two critical components in the \ac{dws} loop from which the LSM1B measurements are derived, are from pre-launch thermal-vacuum tests. The \ac{chu}1 and \ac{oba} temperature sensitivities predict long-term drifts a factor of 5 times smaller than what is observed. The temperature sensitivity of the  \ac{lrp} contributes negligible drift.}
        \label{fig:pitch_tempcos}
    \end{figure*}

Fig. \ref{fig:pitch_tempcos} shows the LSM1B pitch measurements for GF1 against predictions of the long-term drift due to temperature sensitivities in the star camera,  \ac{oba} and \ac{lrp}. The   \ac{oba} and \ac{lrp} are two critical components in the \ac{dws} loop from which the LSM1B angles are derived. We chose not to include \ac{imu} temperature sensitivities in this analysis as the attitude Kalman filter removes low frequencies from the  \ac{imu}  measurement.  The optical bench assembly was built to have a temperature sensitivity $< \unit[5]{\mu rad \text{ K}^{-1}}$ \cite{nick}. The  \ac{lrp}  has a low temperature coefficient ($\unit[5]{nrad \text{ K}^{-1}}$), based on pre-launch measurements in thermal-vacuum tests, which mean the long-term drift from this component is negligible. Fig. \ref{fig:pitch_tempcos} shows both the star camera temperature sensitivity (temperature coefficient $ = \unit[1.4] {\mu rad \text{ K}^{-1}}$ [personal communication John Leif J\o rgensen, DTU, Denmark]) and the  \ac{oba}  temperature sensitivity are the closest to explaining the drift, however even these fall a factor of 5 short. Although the temperature sensitivities of neither the star cameras or \ac{dws} loop are currently able to explain the observed variations, we reinforce that this has not been measured directly in-orbit but is based on pre-launch measurements. The pre-launch temperature sensitivity measurements for all attitude sensors can be found in Table \ref{table2}.

An alternative explanation for the long-term drift is flexing of the satellite body. As Fig. \ref{fig:spacecraft_design} shows, the star camera,  \ac{imu}  and laser beam steering components are mounted at different points, spread out across the satellite body. Any flexing of the satellite would then affect these attitude sensors differently.
	\begin{table}[htbp!] 
			\caption{Measurement noise and thermal sensitivities of the star cameras,  \ac{imu}  and  \ac{dws}  sensors of GRACE-FO.  The  \ac{imu}  thermal sensitivities are not included because the attitude Kalman filter removes low frequencies from the  \ac{imu} measurements.}
								\begin{center}
									\begin{tabular}{|l|l|l|}
										\hline
										\textbf{Sensors} & \textbf{Measurement noise}  & \textbf{Thermal sensitivity} \\
										& ($\unit[]{\mu rad\text{ Hz}^{-1/2}}$) &($\unit[]{\mu rad\text{ K}^{-1}}$)\\
										\hline \\
										Star cameras &   $8.5/\sqrt{\text{f}/\text{Hz}}$ &  $\unit[1.4]{}$  \\
										\hline
										 \ac{imu}  & 0.105/(f/Hz)  &  -  \\
										 \hline
										 \ac{dws}  & Pitch: 1.29  &   \ac{lrp} : $\unit[5\times 10^{-3}]{}$ \\ 
       											 & Yaw: 1.83   &    \ac{oba} : $<\unit[5]{}$ \\
											 \hline
									\end{tabular}
								\end{center}\label{table2}
		\end{table}

 %
%~~~~~~~~~~~~~~~~~~~~~~~~~~~~~`
\subsection{Characteristics of angles in the orbit and time domain}\label{sect:aol_sca_lsm}%
\begin{figure*}[htbp!]
        \centering
        \includegraphics[width=0.96\textwidth, trim={6 0 20 8}, clip]{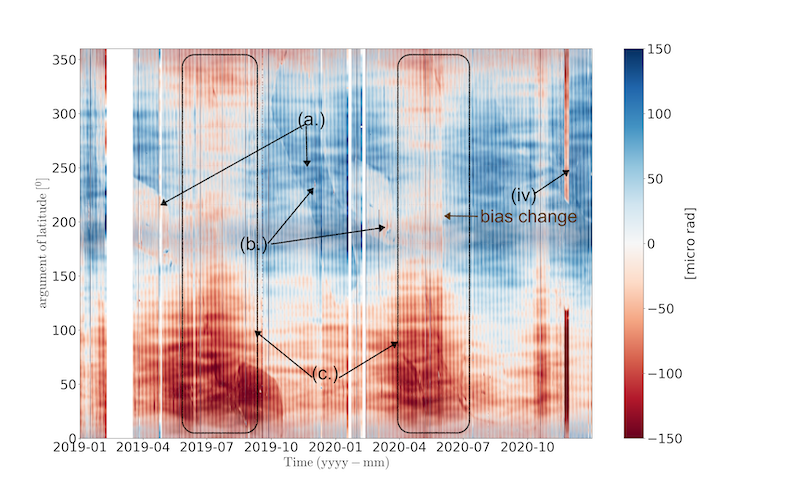}   \rotatebox{90}{\hspace{4cm}LSM1B} \\ 
         \includegraphics[width=0.96\textwidth, trim={6 0 20 8}, clip]{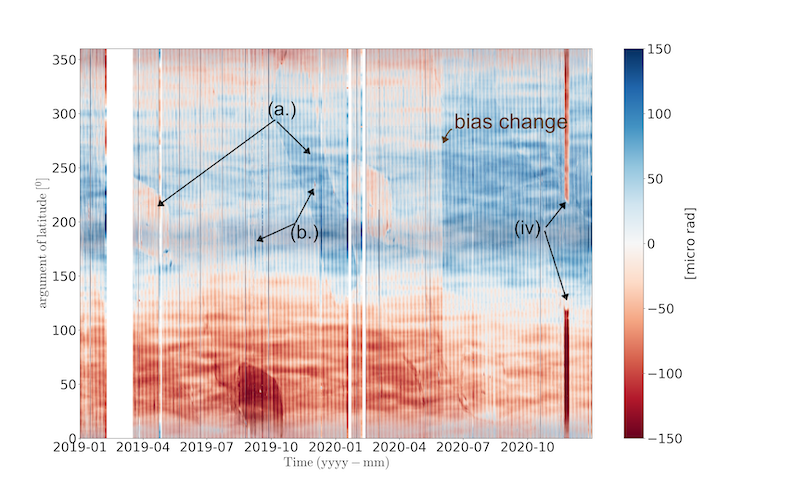}  \rotatebox{90}{\hspace{5cm}SCA1B$_\text{LOS}$} 
        \caption{Pitch angle variations of GF1 as seen in LSM1B (top panel) and SCA1B$_\text{\ac{los}}$ (bottom panel) datasets during the year 2019 and 2020. Features marked as `(a.)'\&`(b.)' are related to the Sun and Moon blindings; `(c.)'  are the temperature dependent variations in the LSM1B pitch angles; `(iv.)' are the disturbances due to thruster calibration tests performed from November 16 to 25, 2020.}
        \label{fig:pitch_gf1_scalsm}
    \end{figure*}

    Here we investigate the differences in pitch and yaw angles of SCA1B$_\text{ \ac{los}}$ and LSM1B data along the argument of latitude and time. 
    The argument of latitude is defined as the angle between the ascending node and the satellite at an epoch \cite{montenbruck2002}.   In Fig. \ref{fig:pitch_gf1_scalsm}, the LSM1B pitch angles are plotted along the argument of latitude (on the y-axis) and time (on the x-axis). Here, the y-axis varies between $0^{0}$ to $360^{0}$ which starts from equator $0^{0}$, ascends to the north pole $90^{0}$ further descending through the equator reaches south pole $270^{0}$ and completes an orbit by reaching again at equator $360^{0}$.  
    The plot of observations along the argument of latitude and time helps us in analyzing their systematic characteristics. We investigate the angle variations using such plots to identify the systematics which are not revealed by the time-series plots.

 \begin{figure*}[htbp!]
\centering
\subfloat[Differences in between the  GF1 pitch angles of SCA1B$_\text{\ac{los}}$ and LSM1B plotted along the argument of latitude.]{
	 \label{fig:diff_scaimu}
	       \includegraphics[width=0.97\textwidth, trim={6 0 20 8}, clip]{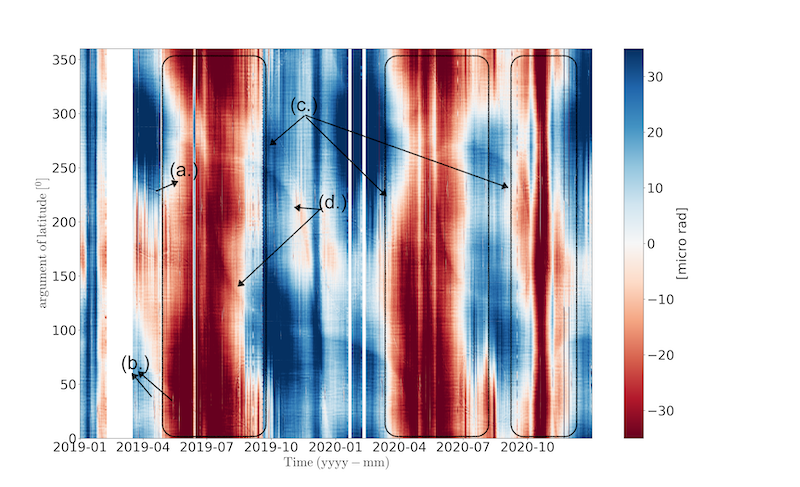} } 
 
\subfloat[Availability of star camera heads on GF1 and spacecraft shadow transitions.]{
	 \label{fig:gf1_scaIDs}
	  \includegraphics[width=0.97\textwidth, trim={6 0 17 0}, clip]{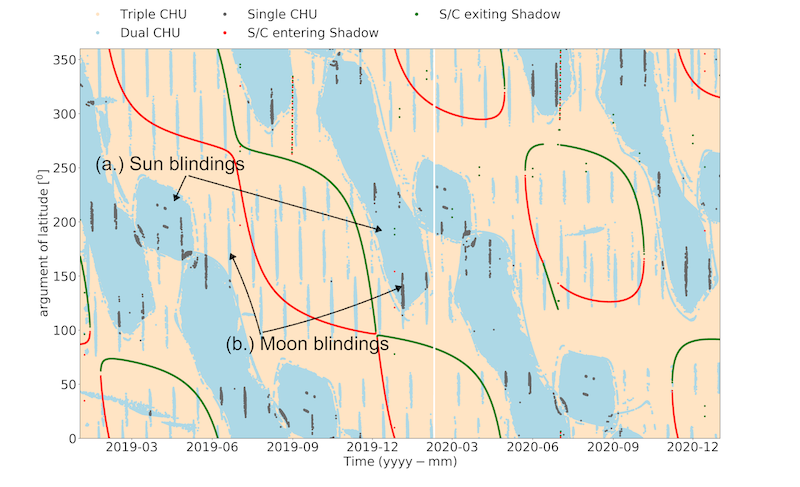} } 
\caption{In Fig.\ref{fig:diff_scaimu} features marked as `(a.)' and `(b.)' shows Sun and Moon blindings; `(c.)' shows differences due to the temperature dependent changes in pitch angle; `(d.)' shows the spacecraft shadow transitions. Sun and Moon blindings into Star Camera Head Units and spacecraft shadow transitions are  shown in Fig. \ref{fig:gf1_scaIDs}. }
\label{fig:diffs_scaid}
\end{figure*}

   Fig. \ref{fig:pitch_gf1_scalsm} shows  GF1 pitch angles of LSM1B (top panel) and SCA1B$_\text{\ac{los}}$  (bottom panel) datasets during the year 2019 and 2020. Both set of pitch angles, although coming from the independent sensors, show attitude variations in between  similar range of magnitude. Similar attitude characteristics are seen in both sets of angles, for example, characteristics related to the periodic star camera Sun and Moon blindings are prominent in the angles (cf. Figs.\ref{fig:pitch_gf1_scalsm} \& \ref{fig:gf1_scaIDs}). The period of availability of star camera heads along with the spacecraft shadow transitions is shown in Fig. \ref{fig:gf1_scaIDs}. 
    Due to the anisotropic noise characteristics of star cameras, which had already been investigated by \cite{bandi2012} for GRACE mission, we see the attitude of the spacecraft affected by the availability of star cameras. 
      
      Besides the Sun and Moon blinding related characteristics in LSM1B pitch angles,  the large variations from mid-May until mid-August 2019, April to June 2020 and in October 2020  are due to the temperature correlated dip as also seen in Fig. \ref{fig:py_lrp_gf1}. These characteristics and their possible explanations has already been discussed in Sec. \ref{sec:temporal}.  
       
    The long term variations seen in Fig. \ref{fig:diff_scaimu} (marked as `(c.)') correlates with temperature, discussed in Sec. \ref{sec:longterm}. 
 Also, slight signatures, especially the boundaries, related to the Sun and Moon blindings into star cameras are visible, also marked as `(a.)' and `(c.)' in Fig. \ref{fig:diff_scaimu} \& \ref{fig:gf1_scaIDs}.  Since the blindings due to Sun and Moon are visible in LSM1B and  SCA1B$_\text{LOS}$  angles (cf. Fig. \ref{fig:pitch_gf1_scalsm}), the related differences (shown in Fig. \ref{fig:diff_scaimu}) are from both set of angles.  Besides those variations, the systematics related to the satellite entering and exiting Earth shadow are also visible in the differences plotted in Fig. \ref{fig:diff_scaimu} (marked as `(d.)'). Spacecraft shadow transitions are shown in Fig. \ref{fig:gf1_scaIDs}. At the moment, we do not fully understand: how do the spacecraft Earth shadow transitions affect the attitude, which dataset (LSM1B/SCA1B) is affected by them and to what extent.  \newline
 As we see similar characteristics in the differences of GF1 yaw and GF2 pitch and yaw angles as well, we limit our discussion to the differences in pitch angles of GF1 only.
\subsection{Spectral differences in between SCA1B$_\text{\ac{los}}$ and LSM1B pointing  angles}\label{sec:spectral}
\begin{figure*}[htbp!]
        \centering
  	\includegraphics[width=0.39\textwidth, trim={10 0 95 8}, clip ]{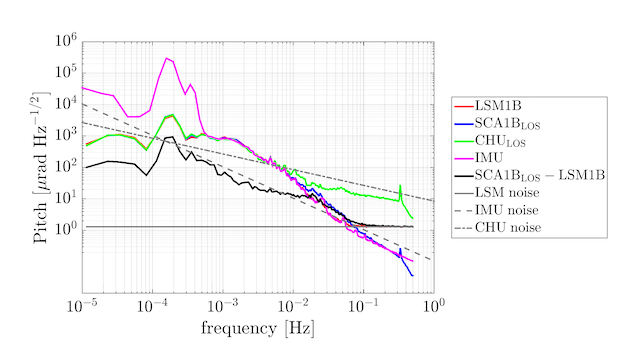} 
 	\includegraphics[width=0.55\textwidth, trim={10 0 8 8}, clip]{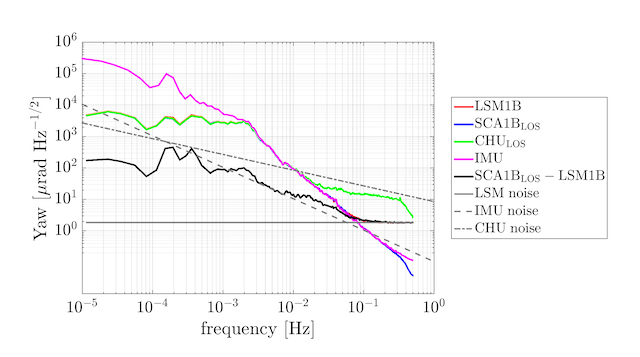}
        \caption{Comparison of the rPSDs of LSM1B,  SCA1B$_\text{LOS}$,  \ac{chu}$_\text{\ac{los}}$ and \ac{imu} angles. The differences between SCA1B$_\text{\ac{los}}$ and LSM1B are also shown along with the noise spectra of  fast steering mirror,  \ac{chu}  and \ac{imu} sensors. Angles are plotted for GF1 for the day January 01, 2019.}
        \label{fig:psd_scafsm_py}
\end{figure*}

Comparing the SCA1B$_\text{\ac{los}}$ and LSM1B angles in the frequency spectrum reveals the two measurements agree over a wide range of frequencies. The once-per-rev error that dominated the time-domain differences between the LSM1B and SCA1B$_\text{\ac{los}}$   GF1 pitch measurements (cf.  Fig. \ref{py_cd_diff}) is prominent in the pitch difference spectrum (Fig. \ref{fig:psd_scafsm_py}). The yaw difference spectrum shows  once and twice-per-rev peaks with comparable power.
However, a few details that weren't apparent in the time-domain analysis become visible: between $\unit[10 - 50]{mHz}$, the LSM1B pitch measurement has lower noise than the SCA1B$_\text{LOS}$ measurement. Additionally, a white frequency noise floor is visible in the LSM1B measurements above $\unit[50]{mHz}$.	

In order to understand the lower noise in LSM1B pitch measurements, we investigated the spectrum of integrated angular rates from IMU measurements, labeled `\ac{imu}’ in Fig. \ref{fig:psd_scafsm_py}, and the spectrum of inter-satellite pointing angles computed from the star camera only combined attitude product, labeled `CHU’.

\ac{chu} product is estimated in the Kalman filter by combining the data from the available star camera head units at a certain epoch. We computed the inter-satellite pointing angles in science reference frame wrt \ac{los} which we represent as \ac{chu}$_\text{\ac{los}}$.

We have also plotted the noise characteristics of \ac{imu}, star camera  and \ac{dws} in Fig. \ref{fig:psd_scafsm_py}. These noise characteristics are summarized in Table \ref{table2}.

The \ac{imu} is the Astrix 120  3-axis fiber-optic gyroscope \cite{airbus}.  It has a specified angle random walk noise equal to
\begin{align}
 	\label{eq:imu_noise0}  \sigma_{\dot{\theta}} &=& 0.0016 \;\text{degree/h/} \sqrt{{\tau/\text{h}}}& =&  A/\sqrt{\tau}  \; \text{rad}\;{\text{s}}^{-1} 	
\end{align}
with  $A = 4.7\times{10}^{-7} \;\text{rad}\; \text{s}^{-1/2}$. $ \sigma_{\dot{\theta}}$  is the Allan deviation of the angle rate. The corresponding angle rate root Power Spectral Density (rPSD) is, after \cite{rutman}, the white spectrum
\begin{align}
	\label{eq:imu_noise1} {\widetilde{\dot{\theta}}} & = \sqrt{2} \;A \;\; \text{rad}\;\text{s}^{-1}\;\text{Hz}^{-1/2}
\end{align}
The angle rPSD follows from the rate rPSD according to
\begin{align}
	 \label{eq:imu_noise2}{\widetilde{\theta}} (f) & =  {\widetilde{\dot{\theta}}}(f)/{(2\pi f)}
\end{align}
giving 
\begin{align}
	\label{eq:imu_noise3} {\widetilde{\theta}} (f) & = 1.05\times10^{-7} \; \text{rad}\;\text{Hz}^{1/2}\;f^{-1}
\end{align}
This  $1/f$ spectrum is shown as the dashed line labeled as IMU noise in Fig. \ref{fig:psd_scafsm_py}.

To model the noise in the star camera only solution, \ac{chu}, we used the Allan deviation for a 3 camera solution, provided by John Leif J\o rgensen, DTU, Denmark [personal communication]. The Allan deviation:
                      \begin{align}
                            \label{eq:chu_noise0}   \sigma_{\text{\ac{chu}}}(\tau) &= 10\; \mu \text{rad}
                       \end{align}
 
Following \cite{rutman}, we estimate the star camera only solution, CHU, has a noise floor with rPSD:

	\begin{align}
		\label{eq:chu_noise1}  \tilde{\theta}_{\text{\ac{chu}}}(f) &= \frac{8.5}{\sqrt{f}}\;  \mu \text{rad} \;{\text{Hz}}^{-1/2}
	\end{align}

As seen in Fig. \ref{fig:psd_scafsm_py}, \ac{imu} sensor has $1/f$ noise floor (\ref{eq:imu_noise3}), \ac{imu} measurements contain significantly less noise in frequencies $\gtrapprox\unit[5]{mHz}$ along pitch axis. Thus, the attitude estimated from the combination of star cameras and \ac{imu}, that is, SCA1B should have reduced high frequency noise due to \ac{imu}. However, in frequencies $\unit[10-50]{mHz}$, the noise in SCA1B$_\text{\ac{los}}$ pitch angle is slightly high than the \ac{imu}. The comparison between SCA1B and the independent LSM1B attitude measurements shows there is potential to further fine-tune the star camera and  \ac{imu} data fusion, to reduce noise in the frequencies between $\unit[10-50]{mHz}$.

As another option, \ac{dws} data can also be combined with the SCA1B/\ac{chu} dataset. We expect that the combination would also be helpful in improving the precision of the pitch angles in frequency range $\unit[10-50]{mHz}$.	

The white LSM noise in Fig. \ref{fig:psd_scafsm_py} is approximately equal to the broadband noise of the \ac{pss} sensor \cite{kaman5100} and is roughly a factor of 3 larger than the analog-to digital converter quantization error. From the data, we estimate the spectrum of the white noise in the pitch and yaw LSM1B measurements to be:
	
    \begin{align}
        \label{eq:pitch_quant} \tilde{\theta}_{\text{LSM,Pitch}} & =  1.29\; \unit[]{\mu \text{rad}\; {Hz}^{-1/2}}\\
        \label{eq:yaw_quant} \tilde{\delta}_{\text{LSM,Yaw}} & = 1.83\; \unit[]{\mu \text{rad} \; {Hz}^{-1/2}}.
    \end{align}
In Equations \eqref{eq:pitch_quant} \& \eqref{eq:yaw_quant}, there is a $\sqrt{2}$ difference between the quantization noise of pitch and yaw because the laser reflects off the mirror at $\unit[45]{^{\circ}}$  and this results in a difference in in-plane (yaw) and out-of-plane (pitch) reflections.
  %  \newpage
%~~~~~~~~~~~~~~~~~~~~~~
    \section{Discussions and Outlook}\label{sec_outlook}
    	In this paper, we presented the first analysis of \ac{lri} \ac{dws}  data which provides the pointing along \ac{los}.  The active beam steering uses \ac{fsm} which is commanded to maintain the inter-satellite laser link between the two spacecraft. From the \ac{lri} telemetry, we use sensed orientation of mirror to compute \ac{los} pitch and yaw angles (Sec. \ref{sec_1a21b}).
		The \ac{dws} provided \ac{los}  pitch and yaw angles is an additional measurement of relative spacecraft attitude. 
	
	We presented the LSM1B data characteristics along with its comparison to SCA1B data. SCA1B data is  used in computing the gravity field solutions.
	The comparison of LSM1B  and SCA1B$_\text{\ac{los}}$  angles  show their difference was varying  up to $\unit[40]{\mu rad}$ peak-to-peak throughout the year 2019 and 2020 (Sec. \ref{sec:compare_dws_sca}). However these differences are at the required level of precision with per-rev systematics in them. 

	 Our analyses show that the LSM1B pitch angles are moderately correlated to the spacecraft temperatures as opposed to the LSM1B yaw and SCA1B$_\text{\ac{los}}$ angles. 
	Candidate explanations:	 \begin{enumerate}
	 	\item The small correlations  between the SCA1B$_\text{\ac{los}}$ and temperatures could be due to SCA1B product being computed in a Kalman filter where the noise parameters are estimated as part of processing. This could reduce such systematics in the product.  
		\item Another reason could be, since star cameras and \ac{imu} are used to correct the attitude, any long-term temperature sensitivity in these sensors would be detected by the \ac{aocs} loop onboard and corrected for. 
		\item Although the low frequency drift is larger in the LSM1B angles than it is in the SCA1B$_\text{\ac{los}}$ angles, this doesn't mean the LSM1B isn't measuring a real tilt of the spacecraft. 
	 Since the LSM1B and SCA1B$_\text{\ac{los}}$ attitude sensors are two independent attitude measurements, any flexing of the spacecraft will result in different LSM1B and SCA1B$_\text{\ac{los}}$  measurements.
	 \end{enumerate}

	Angles when plotted along the argument of latitude (Sec. \ref{sect:aol_sca_lsm})  shows similar attitude characteristics which are visible in the time-series, along with the characteristics such as periodic blindings by Sun and Moon (see Fig. \ref{fig:pitch_gf1_scalsm}). Their differences also contain the signatures of periodic blindings and  the signatures related to satellite transitions into and out of the shadow regions, shown in Fig. \ref{fig:diff_scaimu}. Since the blindings affect both LSM1B and SCA1B$_\text{\ac{los}}$ angles, both set of angles are the source of presence of blinding related signatures in their differences.

		The root power spectral density comparison of pitch angles ( cf. Fig. \ref{fig:psd_scafsm_py} in Sec. \ref{sec:spectral}) show that the LSM1B data contain less noise in frequencies between $\unit[10-50]{mHz}$. Subsequent comparison with \ac{imu} measurements and attitude estimated only from star cameras  revealed the SCA1B Kalman filter could be improved. This has demonstrated the benefit of having an independent attitude such as the LSM1B dataset. 
		The fast steering mirror attitude measurements have a white noise floor, empirically found to be $\unit[1.29]{\mu rad\; {Hz}^{-1/2}} $ in pitch and $\unit[1.83]{\mu rad\; {Hz}^{-1/2}}$ in yaw (shown in Fig. \ref{fig:psd_scafsm_py}). As evident from Fig. \ref{fig:psd_scafsm_py}, the noise level of the fast steering mirror measurements is well below the noise of conventional spacecraft-associated orientation sensors, namely star cameras and \ac{imu}. The comparative long-term stability is less evident, as indicated in Fig. \ref{fig:pitch_tempcos}.  We note however that the gravity signal is concentrated in frequencies between 0.2 and 20 mHz \cite{spero}, suggesting that the long-term precision of spacecraft orientation should not directly affect the quality of the recovered gravity field.  Subsequent studies will exploit the low-noise orientation readout of the fast steering mirror signals for potential improvement of the \ac{lri}-derived gravity fields.
 
%%%%%%%%%%%%%%%%%%%%%%
 \section*{Acknowledgment}
 We would like to thank Christopher McCullough for answering our sensor specification related questions.  We would also like to thank the reviewers for their thoughtful comments and efforts towards improving our manuscript. \newline
The research was carried out at the Jet Propulsion Laboratory, California Institute of Technology, under a contract with the National Aeronautics and Space Administration (80NM0018D0004).

\begin{thebibliography}{00}

\bibitem{gfo2019}
R.P.Kornfeld, B.W.Arnold, M.A.Gross, N.T.Dahya \& W.M.Kilpstein,  ``GRACE-FO:  The Gravity Recovery and Climate Experiment Follow-On mission", \emph{Journal of spacecraft and rockets},  2019,\url{https://doi.org/10.2514/1.A34326}.
\bibitem{felix2020}
F.W.Landerer, F.M.Felchtner, H.Save, F.H.Webb, T.Bandikova, W.I.Bertiger \emph{et al.}, ``Extending the global mass change data record: GRACE Follow-On instrument and science data performance", Geophysical Research Letters, 47, e2020GL088306, 2020. \url{https://doi.org/10.1029/2020GL088306}.
\bibitem{tapley2004} 
B.D.Tapley, S.Bettadpur, J.C.Ries, P.F.Thompson \& M.M.Watkins, ``GRACE measurements of mass variability in the Earth System", \emph{Science}, Vol. 305, Issue 5683, pp. 503-505, 2004, doi: \url{https://science.sciencemag.org/content/305/5683/503}.
\bibitem{tapley2019}
B.D.Tapley, M.M.Watkins, F.Flechtner, C.Reigber  \emph{et al.}, ``Contributions of GRACE to understanding the climate change", \emph{Nature climate change}, VOL 9, 358–369,  2019, \url{https://doi.org/10.1038/s41558-019-0456-2}.
\bibitem{sheard2012}
B.S.Sheard, G.Heinzel, K.Danzmann  \emph{et al.}, ``Intersatellite laser ranging instrument for the GRACE Follow-on mission", \emph{Journal of Geodesy}, Vol. 86, 1083-1095, 2012,  \url{https://doi.org/10.1007/s00190-012-0566-3}. 
\bibitem{abich2019}
K.Abich, G.Heinzel, K.McKenzie \emph{et al.}, ``In-Orbit performance of the GRACE Follow-On Laser ranging interferometer", \emph{Phys. Rev. Lett.} 123, 031101, 2019, \url{https://doi.org/10.1103/PhysRevLett.123.031101}
 \bibitem{daniel2014}
D.Schuetze , G.Stede, V.Mueller \emph{et al.}, ``Laser beam steering for GRACE Follow-On inter-satellite interferometry", \emph{Optics Express}, 24117, Vol. 22, No. 20, 2014, \url{https://doi.org/10.1364/OE.22.024117}.
\bibitem{bandi2014} 
T.Bandikova \& J.Flury, ``Improvement of the GRACE star camera data based on the revision of the combination method", Advances in Space Research, Volume 54, Issue 9, Pages 1818-1827,  2014, \url{https://doi.org/10.1016/j.asr.2014.07.004}.
\bibitem{henry2020}
 H.Wegener, V.Mueller, G.Heinzel  \& M.Misfeldt, ``Tilt-to-length coupling in GRACE Follow-On Laser Ranging Interferometer", \emph{Journal of spacecraft and rockets}, Jul. 2020,  \url{https://arc.aiaa.org/doi/10.2514/1.A34790}.
 \bibitem{bandi2018}
 T.Bandikova \& N.Harvey, ``GRACE Follow-On attitude solution", \emph{GRACE/GRACE-FO Science Team Meeting}, October 9-11, 2018, Potsdam Germany.
 \bibitem{harvey2019}
 N.Harvey \& C.Sakumura, ``Results from a GRACE/GRACE-FO attitude reconstruction Kalman filter", \emph{Journal of Geodesy},  Vol. 93, Issue 10, pp 1881–1896, 2019, \url{https://doi.org/10.1007/s00190-019-01289-z}.
\bibitem{bandi2012}
T.Bandikova, J.Flury \& U.D. Ko, ``Characteristics and accuracies of the GRACE inter-satellite pointing", \emph{Advances in Space Research}, Volume 50, Issue 1, Pages 123-135, 2012, \url{https://doi.org/10.1016/j.asr.2012.03.011}
\bibitem{gfoupdate}
GRACE-FO Satellite Switching to Backup Instrument Processing Unit:\url{https://gracefo.jpl.nasa.gov/news/139/grace-fo-satellite-switching-to-backup-instrument-processing-unit/}, last access: 11 September 2020.
\bibitem{save2019} H.Save,  ``GRACE-FO Science Operations Report", \emph{GRACE/GRACE Follow-On Science Team Meeting 2019}, October 8-10, 2019, Pasadena, California, USA.
  
\bibitem{nick} K. Nicklaus, M. Herding, A. Baatzsch, M. Dehne, C. Diekmann, K. Voss, F. Gilles  \emph{et al.},  ``Optical Bench of the laser ranging interferometer on GRACE Follow-On",  In International Conference on Space Optics—ICSO 2014, vol. 10563, p. 105632I. International Society for Optics and Photonics, 2017 \url{https://doi.org/10.1117/12.2304195}.

 \bibitem{montenbruck2002} O.Montenbruck \& E. Gill,  ``Satellite Orbits - Models, Methods and Applications", Springer-Verlag, Berlin Heidelberg, XI, 369, ISBN:978- 3-642-58351-3,  2002, doi:\url{10.1007/978-3-642-58351-3}.

\bibitem{airbus} AIRBUS. ASTRIX inertial measurement units/IRU series, (2021, March 25), Available:\url{https://www.airbus.com/space/spacecraft-equipment/avionics/astrix.html}.

\bibitem{rutman} J. Rutman, ``Characterization of phase and frequency instabilities in precision frequency sources: Fifteen years of progress", \emph{in Proceedings of the IEEE}, Vol. 66, no. 9, pp. 1048-1075, Sept. 1978,  \url{https://ieeexplore.ieee.org/document/1455349}.

\bibitem{kaman5100} KD-5100 Differential Resolution to a Nanometer,  Sensor Data Sheet, \url{https://www.kamansensors.com/product/kd-5100/}, last access: March 30, 2021.

 \bibitem{spero}
Robert  Spero,  ``Point-mass sensitivity of gravimetric satellites", \emph{Advances in Space Research} 67, no. 5,  1656-1664, 2021, \url{https://doi.org/10.1016/j.asr.2020.12.019}.
%%%%%%%%%%%%%%%%%%%%%%%%%%%%
 \end{thebibliography}
\end{document}